\documentclass[useAMS,usenatbib]{mn2e}
\usepackage{graphicx}
\usepackage{txfonts}
\usepackage[authoryear]{natbib}
\bibpunct[,]{(}{)}{;}{a}{,}{,}
\begin{document}

\title[The N/O evolution in galaxies]{The Nitrogen-to-Oxygen evolution
in galaxies: the role of the star formation rate}

\author[Moll\'{a} et al.]
{M.~Moll\'{a},$^{1}$\thanks{E-mail:mercedes.molla@ciemat.es}
J.~M.~V\'{\i}lchez,$^{2}$ M.~Gavil\'{a}n,$^{3}$ and A.~I.~D\'{\i}az$^{3}$\\
$^{1}$Departamento de Investigaci\'{o}n B\'{a}sica, C.I.E.M.A.T,
 Avda. Complutense 22, 28040, Madrid, (Spain) \\
$^{2}$Instituto de Astrof\'{\i}sica de Andaluc\'{\i}a (CSIC), 
Apdo. 3004, 18080 Granada, Spain \\
$^{3}$Departamento de F\'{\i}sica Te\'{o}rica,
Universidad Aut\'onoma de Madrid, 28049 Cantoblanco, Madrid (Spain)}
\date{Accepted Received ; in original form }

\pagerange{\pageref{firstpage}--\pageref{lastpage}} \pubyear{2006}

\maketitle
\label{firstpage}

\begin{abstract}
The main objective of the present work is to ckeck if the star
formation efficiency plays a relevant role in the evolution of the
relative abundance N/O.  In order to explore this idea, we analyze the
evolution of the nitrogen-to-oxygen ratio as predicted by a set of
computed theoretical models. These models consist of simulated
galaxies with different total masses which are evolved assuming
different collapse time scales and different star formation
efficiencies. The combinations of these two parameters produce
different star formation histories, which in turn have, as we show, 
an important impact on the resulting N/O ratio.  Since we want to
check the effect of variations in these efficiencies on our models
results, the same stellar yield sets are used for all of them. The
selected yields have an important contribution of primary nitrogen
proceeding from low and intermediate mass stars, which implies that N is
ejected with a certain delay with respect to O. It allows to obtain,
as we demonstrate, a dispersion of results in the N/O-O/H plane when
star formation efficiencies vary which is in general agreement with
observations.  The model results for the N/O abundance ratio are in
good agreement with most observational data trends.  In particular,
the behavior shown by the extragalactic H{\sc ii} regions is well
reproduced with present time resulting abundances. Furthermore, the
low N/O values estimated for high-redshift objects, such as those
obtained for Damped Lyman Alpha (DLA) galaxies, as well as the higher
(and constant) values of N/O observed for irregular and dwarf galaxies
or halo stars, can be simultaneously obtained with our models at the
same low oxygen abundances $\rm 12+log(O/H)\sim 7$.  We therefore
conclude that, even though there seems to be a general believe that
abundance ratios depend mostly on stellar yields, these are
not the only parameter at work when both elements are ejected by stars
of different mass range, and that differences in the star formation
history of galaxies and regions within them are a key factor to
explain the data in the N/O-O/H plane.

\end{abstract}

\section{Introduction}

The nitrogen abundances have posed an important debate which lasts up
to now. First, because it is created in two types of stars, what rises
the doubt about which proportion is produced in massive stars and how
much N is ejected by low and intermediate mass (LIM) stars. Second,
the possible primary or secondary origin of nitrogen implies some
questions about what is the contribution of each type produced in each
star. The question is then how much N is created in each stellar mass
range and what fraction of it is primary?

From an observational point of view, the need for a primary component
of N appeared very soon.  \citet{edm78} and \citet{allo79} already
concluded, from the analysis of the then available extragalactic H{\sc
ii} region data, that N behaves, at least partially, as a primary
element. This was later supported by N measurements for Galactic
metal-poor halo stars \citep{bar83,tom84,lai85,car87}, and by
additional extragalactic H{\sc ii} region observations
\citep{mcc85,vila93}.  The number of N data points has increased
extraordinarily since then \citep[see][and references therein]{hen00,con02} 
showing the same basic trend: 1) a strong slope
for the N/O data against oxygen abundance for metal-rich regions, a
well explained behaviour by the secondary character expected from the
CNO cycle production, and 2) a flat line for the data of low mass and
irregular galaxies and halo stars with $\rm 12 + \log{(O/H)}<8$. Some
very recent observations of halo stars \citep{isr04,spi05} have
increased the number of data in the low abundance end. They show a
large scatter in N/O at low metallicities, and some high N/O ratios,
implying an important primary N production at very early evolutionary
times.

Since low metallicity galaxies have been considered to be young
objects undergoing their first burst of star formation
\citep{thuan95}, it was assumed that they have had no time for their
LIM stars to evolve and eject the nitrogen produced by them.
Moreover, the dispersion found for their N/O abundances was really
low. This way, a number of authors have claimed that massive stars are
the main primary N producers. There are two facts against this
argument: 1) these galaxies are not so young as it was thought, since
recent observations indicate that they host stellar populations which
are at least $10^{7}-10^{8}$ yr old, reaching even 1 Gyr is some cases
\citep{leg00,tols03,cai03,vzee04,thuan05}; 2) if LIM stars with masses
in the range 4--8 $\rm M_{\odot}$ eject some amount of primary N, this
can be in the interstellar medium (ISM) within a time as short as
$\sim$ [50 -- 200] Myr, since the mean-lifetimes of these stars are in
this range. Moreover, low oxygen abundance do not necessary imply
short time-scales. In fact, we claim that it is essential to change
the usual scheme that identify oxygen abundances with evolutionary
time, a misunderstander concept that lead to erroneous conclusions.

On the other hand, there is no clear mechanism which may produce
primary N in massive stars. Some authors working on stellar evolution
have searched for new possibilities, the most plausible one being the
effect of rotation on stars.  \citet{mey02} have computed some yields
from stellar models including rotation obtaining that intermediate and
massive stars may produce primary N, mostly at very low metallicity.
However, \citet{chia05} have included these yields in a Galactic
chemical evolution (GCE) model finding that an extra-- production of N
in low metallicity massive stars by a large factor, between 40 and 200
along the mass range, is still necessary to explain the data for very
metal-poor halo stars since the yields used do not produce a
sufficient amount of primary N. Recently, \cite{chia06} use new
(and still unpublished) yields for metal-poor massive stars which
rotate rapidly thus producing large amounts of primary nitrogen with
which it is possible to increase the N/O abundance up to similar
levels of the Galactic halo stars. These preliminary yields are still
uncertain since they are computed assuming a non-observed very high
rotation velocity for these low-metallicity stars.
\footnote{In fact, as \cite{hir05} stated, {\sl the ratio
$v_{ini}/v_{c}$ decreases at very low metallicity for the initial
velocity of 300 km $^{-1}$ because stars are more compact at lower
metallicity}.}.

The situation has become more complicated because abundances of N for
DLA galaxies have been derived \citep[][and references
therein]{pet02,pro02,cen03}. When these recent observations are
included in the sample, a large dispersion appears, in evident
disagreement with the hypothesis of all the primary N being produced
by massive stars.  DLA data show lower values of N/O than those,
populating the flat slope, obtained for the irregular and blue compact
galaxies and for the halo stars of similar oxygen abundance. These
high redshift objects have oxygen abundances which correspond to
slightly evolved objects as those observed in the local universe, but
in this case some values of N/O are much smaller than the local ones.
This fact appears incompatible with primary N being produced by
massive stars, as those very metal-poor ones rotating at very
high velocity used in \cite{chia06}, since in that case it results
difficult to conceive which mechanism could produce a different final
yield (and therefore a different point on the N/O-O/H plane) for each
DLA object.

From the theoretical point of view, a secondary production is expected
in most stars, as corresponds to the CNO cycle, while the primary
contribution should arise from intermediate mass stars ($4 \leq
M/M_{\odot} \leq 8$).  These stars may suffer, during the Asymptotic Giant
Branch (AGB) phase, dredge-up episodes and hot bottom burning processes, due
to which some primary N may be created. This fact has been well known for some
decades \citep{rv81,ser86}. In fact, most of yield sets \citep[as][]{vhoek97,
mar98} presently used for the chemical evolution models include a contribution
of primary N produced by these processes in LIM stars. The AGB phase is
still far from being well understood, and there exist large uncertainties
about the exact amount of primary nitrogen produced through the HBB
process. Probably due to that reason, the up to now computed models do not
seem be able to reproduce most of the data.

The idea that a constant N/O may only be obtained with a primary N
from massive stars is indeed an oversimplification which does not take
into account the stellar lifetimes.  \citet{vila93} and \citet{pil03},
by analyzing the N/O values in galaxies of different morphological
type, concluded that a long-time-delayed contribution to the N
production must exist. \citet{vila93} suggested that most data fall in
a region limited by two extreme closed box models defined by distinct
delays, produced by different star formation efficiencies, and/or ages
of galaxies, thus explaining the observed dispersion.

This way, \citet{hen00} and \citet{pran03} --hereinafter HEN and PRAN,
respectively -- computed realistic chemical evolution models using
yields from \citet{vhoek97} for LIM stars, with a contribution of
primary N. Although their models produce a level of N/O higher
than observations for Z low, they show that, actually, a N/O {\sl vs}
O/H relation much flatter than usual may be obtained with their
models, only decreasing the efficiencies of star formation. The
same finding was also obtained by \cite{lar01} using several sets of
yields, all of them with a proportion of primary N ejected by LIM
stars. These results lead to the idea that stellar yields are not the
only driver of the N/O ratio, at least when LIM stars produce a
quantity of primary N, and that most observations in the plot N/O-O/H
might be reproduced if different star formation rates in the different
galaxies or regions are considered

\citet{gav05} and \citet[][hereinafter GAV05 and GAV06,
respectively]{gav06} have recently calculated some chemical evolution
models for the Milky Way Galaxy (MWG), by using different sets of
stellar yields. They demonstrated in those works that a model using
the set of yields for LIM stars from \citet{gav05} joined to the
massive stellar yields from \cite{woo95} is able to reproduce most of
trends of MWG data.  In the cited model, a flat behaviour of N/O {\sl
vs} O/H appears when the star formation rate is low during a long
time, as it occurs in the Galactic halo or in the the outer regions of
the Galactic disc. In this case, since the evolution is slow, a time
longer than 40 Myr is necessary to reach oxygen abundances $12
+log(O/H) \sim 7$, which is sufficient for LIM stars to eject their
products to the ISM. This, combined with low O/H abundances, produces
high N/O values, similar to those observed, for low abundances.  Most
of the data of the Galaxy halo and disc, in particular those of
nitrogen --see their Fig.12--, are successfully reproduced with that
model.  Furthermore, the model results for the radial regions other
than the Solar Neighbourhood suggest that actually the dispersion
observed in MWG data can be a consequence of different star formation
histories (produced by different collapse timescales to form the disc
and different efficiencies to form stars) in the Galactic radial
regions. This supports the mentioned findings from the above cited
authors: different evolutionary tracks and end points in the O/H-N/O
plane are possible simply by increasing or decreasing the star
formation rate, although other parameters probably also play a role.

 The objective of this work is, therefore, to explore if different
star formation efficiencies may produce different tracks in the N/O-O/H
plane and to check, with a wide grid of chemical evolution models, if
the results are acceptable as compared with the observational data
trends. In order to do this task we need a large number of models with
the same basic assumptions about the evolution and where only the star
formation be varied.  Therefore, we analyze the results obtained for
the evolution of nitrogen with the wide grid of models shown in
\citet{mol05}, which simulates 44 galaxies with different total masses
and 10 possible star formation efficiencies. The chosen sets of yields
used for these computations were those from WW for massive stars and
those from GAV05 for LIM stars, which were already calibrated as
explained above.  An advantage of using these yields is that the
primary N proceeds from LIM stars with which the effect of variable
star formation is probably more evident. \footnote{Since the objective
of this work is to explore if different star formation efficiencies
produce different tracks in the N/O-O/H plane we must use the same
ingredients for all our realizations, except the parameter we want to
check. This is similar to the usually applied method when the stellar
yields are the free parameters. In that case using the same
ingredients, except the stellar yields, for some Galactic chemical
evolution (GCE) models, it is possible to determine what set of yields
is the most adequate.}

In Sect.~2 we analyze the N/O-O/H results obtained with simple
chemical evolution models, by describing how elemental abundances of N
and O evolve. In Sect.~3 we summarize the resulting abundances for the
multiphase chemical evolution model with a contribution of primary
nitrogen proceeding from LIM stars variable with Z.  Sect.~4 is
devoted to the results obtained when the same code is applied to a
wide grid of simulated galaxies with different masses and star
formation efficiencies.  Our conclusions are in Sect.5.

\section{The closed box model predictions}

Nitrogen needs a seed of carbon or oxygen to be created from the
CNO cycle. If this seed already exists when the star forms, N will be
produced as a secondary element, what implies that its relative
abundance N/C (or N/O) will increase proportionally to the original
abundance of C or O. If the seed is created in the same star, just
before N, then N is primary, and its abundance will be proportional to
the C or O abundance, that means that N/C or N/O will maintain a
constant value.  Both trends are easily obtained using the classical
and well known {\sl closed box model} (hereinafter CBM)
following which the abundance of an element $i$, $Z_{i}$, is related
on the gas fraction, $\mu=g/M$, or ratio of the gas mass to the
total mass in the region \citep{pag79,tin80}, through the expression: 
$Z_{i}=p_{i}ln\mu^{-1}$, where $p_{i}$ is the stellar
yield, the new production of an element $i$ by a generation of stars.
If nitrogen and oxygen would be both
primary, with yields $p_{N}=ppN$ and $p_{O}=pO$, respectively,
then the relative abundance $\rm Z_{N}/Z_{O}$ is:

\begin{equation}
\frac{Z_{N}}{Z_{O}}=\frac{ppN}{pO}=cte
\end{equation}

It, instead, nitrogen is secondary, its stellar yield $p_{N}=psN$ will
depend on the abundance of the {\sl seed}, in this case Oxygen. Then:
$p_{N}=psN \times Z_{O}$, and therefore:

\begin{equation}
\frac{Z_{N}}{Z_{O}} = \frac{psN}{2 pO}Z_{O}
\end{equation}

\begin{figure}
\resizebox{\hsize}{!}{\includegraphics[angle=-90]{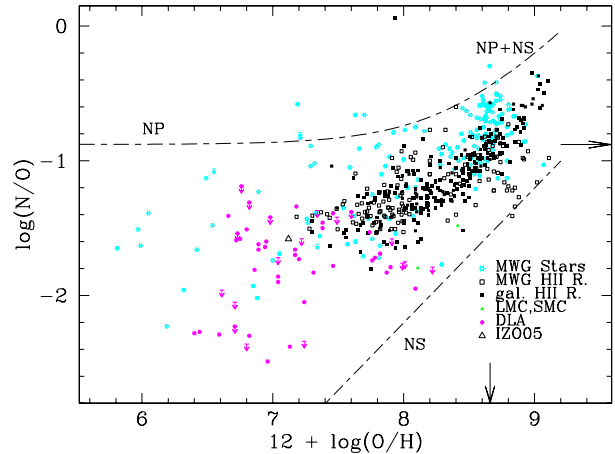}}
\caption{The relative abundance log(N/O) as a function of the oxygen
abundance, $12 + log(O/H)$, when N is secondary, shown
by the line NS, primary, marked as NP, or a combination of
both primary + secondary, marked as NP+NS. The points are the 
available data taken from references of Table~\ref{authors}
for stars of our Galaxy, Galactic and external galaxy H{\sc ii} regions, 
LMC and SMC galaxies and DLA objects as labelled.
The data for the lowest metallicity known galaxy (SBS0335-052) from
\citet{izo05} is marked as an open triangle. The arrows mark the solar
values in the graph and the short-long-dashed lines limit the region
occupied by the observational data.}
\label{no_lim}
\end{figure}

Actually, the mechanism to form N in stars is mostly secondary, with a
behaviour as the one shown by the line named NS in Fig.~\ref{no_lim}.
However, the observed data (taken from references given in
Table~\ref{authors}), indicate, such as we may see in the same figure,
that a primary contribution of N is necessary to explain the behaviour
of data in the Universe. A line as this one marked with NP shows the
expected behaviour with a primary contribution plus a secondary
one. This last one is only apparent when the oxygen abundance is high
enough ($ 12 + log(O/H) > 7.5-8$), marked as NP+NS.  Obviously, if we
change the yield ppN or psN, the lines would move toward higher or
lower N/O abundances. The two lines shown in the figure have, in fact,
been drawn to show the possible range of the data.  It is clear that
only a line cannot explain the observed dispersion but is is also
evident that both components are necessary to obtain the trend shown
by the data.

Nevertheless, a primary behaviour does not imply necessarily a constant
value of N/O for the whole time evolution since we must take into
account the mean-lifetimes of stars. The above results are obtained,
as usual, with the classical hypotheses in the closed box model that
all stellar products are ejected simultaneously in the ISM.  Some
primary nitrogen is easily produced by LIM stars in the range of
[4--8] $M_{\odot}$ \citep{rv81,vhoek97,mar98,gav05} during the Hot Bottom
Burning (HBB) process. Since these stars died after the bulk of oxygen created
by massive stars was ejected, if they are the main producers of
primary N, this N would appear in the ISM later than the oxygen does.  In
this scheme, if the necessary NP was ejected by the LIM stars, one
would expect that N/O will be flat only after the oxygen has already
reached a certain level of abundance, giving time to nitrogen to appear,
but it will not be constant before that moment.

\begin{flushleft}
\begin{table}
\caption{References for N and O stellar abundances used for the comparison}
\label{authors}
\begin{tabular}{l}
\hline
\noalign{\smallskip}
\multicolumn{1}{c}{MWG Stars}\\
\noalign{\smallskip}
\hline
\noalign{\smallskip}
\cite{cle81}               \\
\cite{daf04}               \\
\cite{gra00}               \\
\cite{gum98}               \\
\cite{isr04}(ISR)          \\
\cite{sma01}               \\
\cite{spi05}(SPI)          \\
\hline
\noalign{\smallskip}
\multicolumn{1}{c}{MWG H{\sc ii} Regions}\\
\noalign{\smallskip}
\hline
\cite{car05}               \\
\cite{est99a,est99b,est99c}  \\
\cite{fich91}              \\
\cite{pei79}               \\
\cite{sha83}               \\
\cite{tsa03}               \\
\cite{vil96}               \\
\hline
\noalign{\smallskip}
\multicolumn{1}{c}{External Galaxies H{\sc ii} Regions}\\
\noalign{\smallskip}
\hline
\cite{gar95,gar99}        \\
\cite{izo99}              \\
\cite{izo05}              \\
\cite{nav05}              \\
\cite{tsa03}              \\
\cite{vzee98,vzee06}      \\
\cite{vzee06a}            \\
\hline
\noalign{\smallskip}
\multicolumn{1}{c}{Damped Lyman Alpha Objects}\\
\noalign{\smallskip}
\hline
\cite{cen03}               \\
\cite{pro02}               \\
\cite{pet02}               \\
\hline
\noalign{\smallskip}
\end{tabular}
\end{table}
\end{flushleft}
\normalsize

In that case the evolution of the N abundance is given by 
the following equation, taken from \citet[][Ap.B]{hen00}:

\begin{equation}
Z_{N}=\frac{ppN}{pO}exp(\frac{Z_{\tau}}{pO})(Z-Z_{\tau})*F+\frac{psN
ppC}{2 pO^{2}}Z^{2}+\frac{psN psC}{6 pO^{2}}Z^{3},
\end{equation}

where $ppC$ and $psC$ refer to the carbon primary and secondary
yields, analog to $ppN$ and $psN$, and $F$ is a step function that has
a value of zero except if $Z > Z_{\tau}$. The abundance of N is, this
way, computed as a function of the oxygen abundance, $\rm Z=Z_{O}$,
which will be ejected by short mean-lifetimes massive stars, that
is almost immediately after the formation of the first stars. The
nitrogen has a secondary component that also appears in the ISM from
early times (second and third terms), and other primary contribution,
only produced by LIM stars, which will appear in the ISM with a
certain delay (first term).

\begin{figure}
\resizebox{\hsize}{!}{\includegraphics[angle=0]{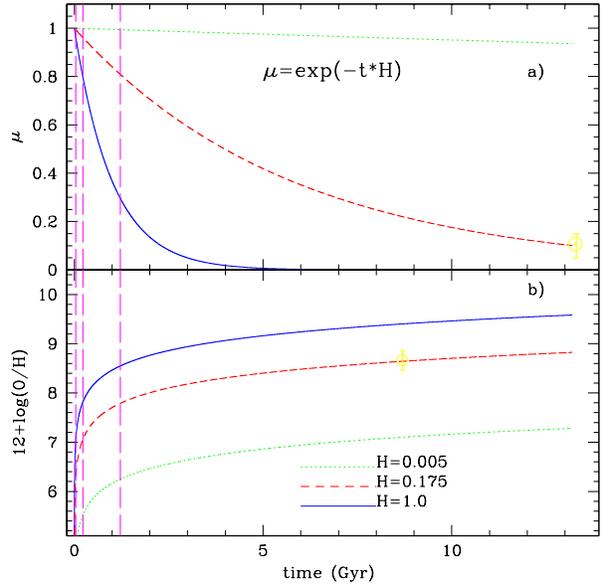}}
\caption{a) The time evolution of the fraction of gas $\mu=exp(-tH)$ 
for three different values of H$=0.005$, 0.175, and 1 as given in
the bottom panel; b) The time evolution of the oxygen abundance $12 +
log(O/H)$ for the same models than a). The long-dashed (magenta) lines mark the
times $t= 0.040$. 0.221 and 1.21 Gyr. The Solar region values are indicated 
in both panels.}
\label{cbm}
\end{figure}

This delay translates in that the primary N will appear only when the
oxygen abundance has already reached a certain value $Z_{\tau}$,
reached in the time when the LIM stars died.  Actually, the LIM stars
have not large mean-lifetimes $\tau$.  Using the age-stellar mass
relation obtained by the Geneva group isochrones \citep{scha92}, stars
of 2, 4, and 8 $M_{\odot}$, have mean-lifetimes of $\tau \sim 1.21$,
0.228, and 0.040 Gyr, respectively, short in chemical evolution
terms. In order to calculate the abundances $Z_{\tau}$ reached when
these stars die, we use again the CBM: $Z_{\tau}=p
\ln{\mu^{-1}(t=\tau)}$.  For computing the values of the gas fraction,
at those times, we assume that it is a decreasing exponential function of time:
$\mu(t)=exp(-t H)$, where H is the efficiency to form stars (equal to the
inverse of the time-scale for consuming the gas, 1/$\tau_{gc}$).  We
show in Fig.~\ref{cbm}a the time evolution of this function $\mu(t)$
for 3 different values of $\rm H=0.005$, 0.175, and 1.0 (or
$\tau_{gc}$=200, 5.75, and 1 Gyr). In panel b) we show the time
evolution of the oxygen abundance, as $\rm 12 +log(O/H)$,
corresponding to each one of these parameters H.  We give in
Table~\ref{ztau} the values of $Z_{\tau}$ reached in these closed
models when $t=\tau$ for stars of 2, 4, and 8 M$_{\odot}$ computed
using $pO=3.3E-3$, $ppC=1.2E-3 $, $psC=0.9 $, $ppN=2.2E-4$ and $psN=0.130$.
as stellar yields for Oxygen, primary and secondary Carbon, and
primary and secondary Nitrogen, respectively.

\begin{table}
\caption{Values of oxygen abundances $Z_{\tau}$ reached when LIM stars
die for different values of the parameter H.}
\label{ztau}
\begin{tabular}{ccccc}
\hline
M$\rm _{star}$ & H$_{1}$ & H$_{2}$ & H$_{3}$  \\
M$_{\odot}$ & 1.000 & 0.175 &  0.005 \\
\noalign{\smallskip}
\hline
\noalign{\smallskip}
2 & 0.399e-2     &  0.695e-3    &  0.200e-4     \\
4 & 0.729e-3     &  0.131e-3    &  0.376e-5     \\
8 & 0.123e-3     &  0.229e-4    &  0.660e-6     \\
\hline
\end{tabular}
\end{table}
\normalsize

\begin{figure}
\resizebox{\hsize}{!}{\includegraphics[angle=0]{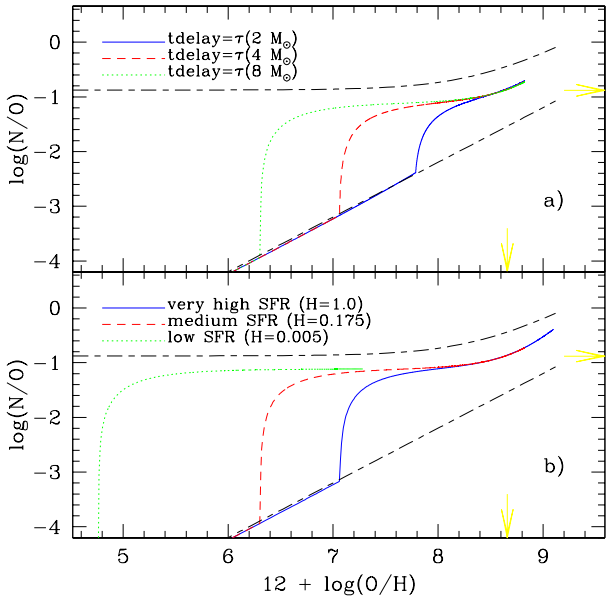}}
\caption{The relation of log(N/O) with the oxygen abundance $12
+log(O/H)$, by assuming that the gas fraction is an exponential
decreasing function of time, obtained by Eq.(4) with two
contributions. The primary component appears in the ISM: a) when stars
of 2, 4 and 8 M$_{\odot}$ die, dotted (green), short-dashed (red) and
solid (blue) lines, respectively, with a parameter H$=0.175$; b) when
stars of 8 M$_{\odot}$ die by assuming different values for H as
marked in the figure.  The short-long-dashed lines and the (yellow)
arrows have the same meaning than in Fig.~\ref{no_lim}.}
\label{evolN}
\end{figure}

In Fig.~\ref{evolN}a) we show how the N/O evolves when O/H increases
in the model corresponding to $H=0.175$ for the three possible values
of column 3 of Table~\ref{ztau}, which would be the oxygen
abundance reached by assuming a primary N proceeding from stars of 2,
4 and 8 M$_{\odot}$. We may distinguish three phases in that figure:

\begin{enumerate}
\item The N/O increases smoothly as secondary from the first phases of
the evolution, for $Z< Z_{\tau}$, with a clear slope that corresponds
to the secondary contribution of massive stars.
\item The NP appears in the ISM when $Z= Z_{\tau}$, increasing very
rapidly the relative abundance (N/O). The final value of this phase is
given by $\frac{ppN}{pO}$. This feature must not be erroneously taken
as a secondary behaviour.
\item After this exponential function, the secondary behaviour
proceeding from LIM stars \footnote{We have use the same psN for both
massive and LIM stars although it may have a different value for each
mass range.} appears again.
\end{enumerate}

Two evident facts appear in that graph:

\begin{itemize}
\item [-] The change from the NS to the NP regime occurs very abruptly.
\item [-] The oxygen abundance at which this change occurs is very
dependent on the star formation rate and on the mass of the first
stars ejecting primary nitrogen.
\end{itemize}

Obviously, if NP would be ejected by stars of 2 $\rm M_{\odot}$,
which is not a realistic case, there would be more time for oxygen
abundances to increase, and so the value $\rm Z_{\tau}$ would  be higher
than in the case of NP ejected by stars of 8 $\rm M_{\odot}$.  In the
last case (8 $\rm M_{\odot}$), the abrupt increase of N/O appears at
the left in the figure N/O --O/H, while the first one ($2M_{\odot}$)
produces this increase at the right of the figure. The higher the mass
of the stars ejecting NP, the lower the oxygen abundance for which the
change takes place.

On the other hand, it is also evident that these $\rm Z_{\tau}$
abundances, shown in Table~\ref{ztau}, depend on the star formation
rate: in Fig.~\ref{evolN}b) we show the results obtained using the
values $Z_{\tau}$ shown in the last line of Table~\ref{ztau}, assuming
that stars of 8 $\rm M_{\odot}$ are the responsible in producing
NP. If the star formation rate is strong ($H=1.0$, solid line) the
oxygen abundance increases very rapidly, so when the primary nitrogen
begins to be ejected, O/H has already reached a high value.  On the
contrary, if the star formation rate is low ($H=0.005$, dotted line),
Oxygen maintains a low abundance for a long time. It is, therefore,
possible that Nitrogen to be ejected when O/H is still below
$12+log(O/H) < 7$.

The closed box model provides approximated results, even for oxygen
abundances, since it is valid only if $Z << 1$ which is not longer
valid when the fraction of gas is small. Furthermore, the hypothesis
of a gas fraction decreasing with time is not valid for a disc which
forms from the gas in-falling from the halo.  In fact, the evolution
of the gas fraction predicted by a CBM with $H=0.175$, which might
represent the Solar vicinity, is very similar to the one obtained with
a numerical model by GAV05 for the Solar region. Both models differs,
however, at the earliest phases: first, the numerical model maintains
a higher value than the equivalent CBM, and then it rests below this
last one. Finally, both reach a similar value of $\sim 0.10$ as
observed.  The oxygen abundances are not equivalent, too, to the CBM
ones. The GAV05 model lasts more time to reach to an oxygen abundance
$12 + log (O/H)=7$. This occurs just when the first LIM stars begin to
die, so it is important to remind that, in a realistic case, when LIM
stars begin to die, the oxygen abundance may be even lower than
predicted by a CBM model.
 
In summary, the stellar yields and the mass range of stars (or their
mean-lifetimes) that produce nitrogen are important to determine the
possible evolution of N/O {\sl vs} O/H but the star formation
history results also essential to determine what is the track followed
by a given region or galaxy in the plane N/O-O/H.

\section{The multiphase chemical evolution model}

\subsection{Abundance calculations}

Now we use a numerical code which takes into account the stellar
mean-lifetimes and other more realistic assumptions for the scenario
and input prescriptions (infall rate, masses, geometry, etc). 
The next calculations have been performed with the
multiphase model, a numerical code developed by \citet{fer92}, and
widely described in that work and in the following ones
\citep{fer94,mol96} of the same group.

As it is explained in \citet{fer92}, the equation to compute the
abundances of each element $i$ in the gas, $X_{i}$, is:

\begin{equation}
\frac{dX_{i}g_{D}}{dt}=-X_{i}\Psi+X_{i,H} f g_{H}+ W_{i}(t)
\end{equation}

$W_{i}$ being the ejected mass rate ({\sl new + old }) for each
element $i$ by the stars, $g_{D}$ the gas mass, $\Psi$ the star
formation rate in the modeled disc region and $f g_{h}$ the infall
rate, assuming as usual that there exists a gas {\sl infall} from the
halo to the disc. The ejected metals mass rates or restitution rates
$W_{i}(t)$ are computed following the formalism of the Q's matrices,
well described in the cited work \citep{fer92} and also in other
works, as \cite{pcb98}. We recommend these works to the reader for a
clear explanations of equations.

The expression to compute the term $W_{i}$ is:

\begin{equation}
W_{i}(t)= \int_{m,\tau=t}^{mup} \sum_{j}{\tilde{Q}}_{i,j}(m) X_{j}(t-\tau(m)) \Psi(t-\tau(m))dm 
\end{equation}

where the expression
$\tilde{Q}_{i,j}={Q}_{i,j} \Phi(m) \Phi(m)$ represent the
matrix $Q_{i,j}$ multiplied by the initial mass function $\Phi(m)$, and $Q_{i,j}$
are defined as:

\begin{equation}
Q_{i,j}(m)=\frac{mexp_{i,j}}{m_{j}}=\frac{mexp_{i,j}}{m X_{j}}
\end{equation}

${mexp_{i,j}}$ being the ejected mass of each element $i$ which was
originally in the star as element $j$.  This way, the total fraction
of element $i$ ejected by each type of stars with mass $m$ is:

\begin{equation}
\frac{mexp_{i}(m)}{m} \Phi(m)=\sum_{j}\frac{mexp_{i,j}}{m}\Phi(m)
=\sum_{j}{\tilde{Q}_{i,j}(m) X_{j}}
\end{equation}

When there are two components primary and secondary for an element, as
it occurs with N, it is necessary to know both contributions for each
star and to compute them separately in the model. The multiphase model
uses for that the equations shown in \cite{fer92}, similar to those
shown by \cite{pcb98} in their Appendix B.  

\subsection{Parametric study}

The above cited expressions are those included in the multiphase
chemical evolution code from \cite{fer92,fer94} where yields from
\citet[][RV]{rv81} for LIM stars and from \citet[][WW86]{woo86} for
massive stars were used to model the Galactic chemical evolution.
Taking into account our previous findings, it is evident that not only
the stellar yields play a role in the resulting evolution of N in a
region or galaxy. The star formation history is very important in
determining the track followed by a given object in the plane
N/O-O/H. A distinct feature is shown in this plane when the primary
nitrogen is ejected to the ISM; and this feature appears at higher or
lower oxygen abundances depending on the star formation rate: the
stronger the star formation rate, the highest the oxygen abundance at
which this characteristic increase appears. Therefore, it is necessary
to analyze how N evolves when different input parameters, related to
the star formation rate, are chosen for a given multiphase chemical
evolution model as this one used in \cite{fer94}.

\begin{figure*}
\resizebox{\hsize}{!}{\includegraphics[angle=-90]{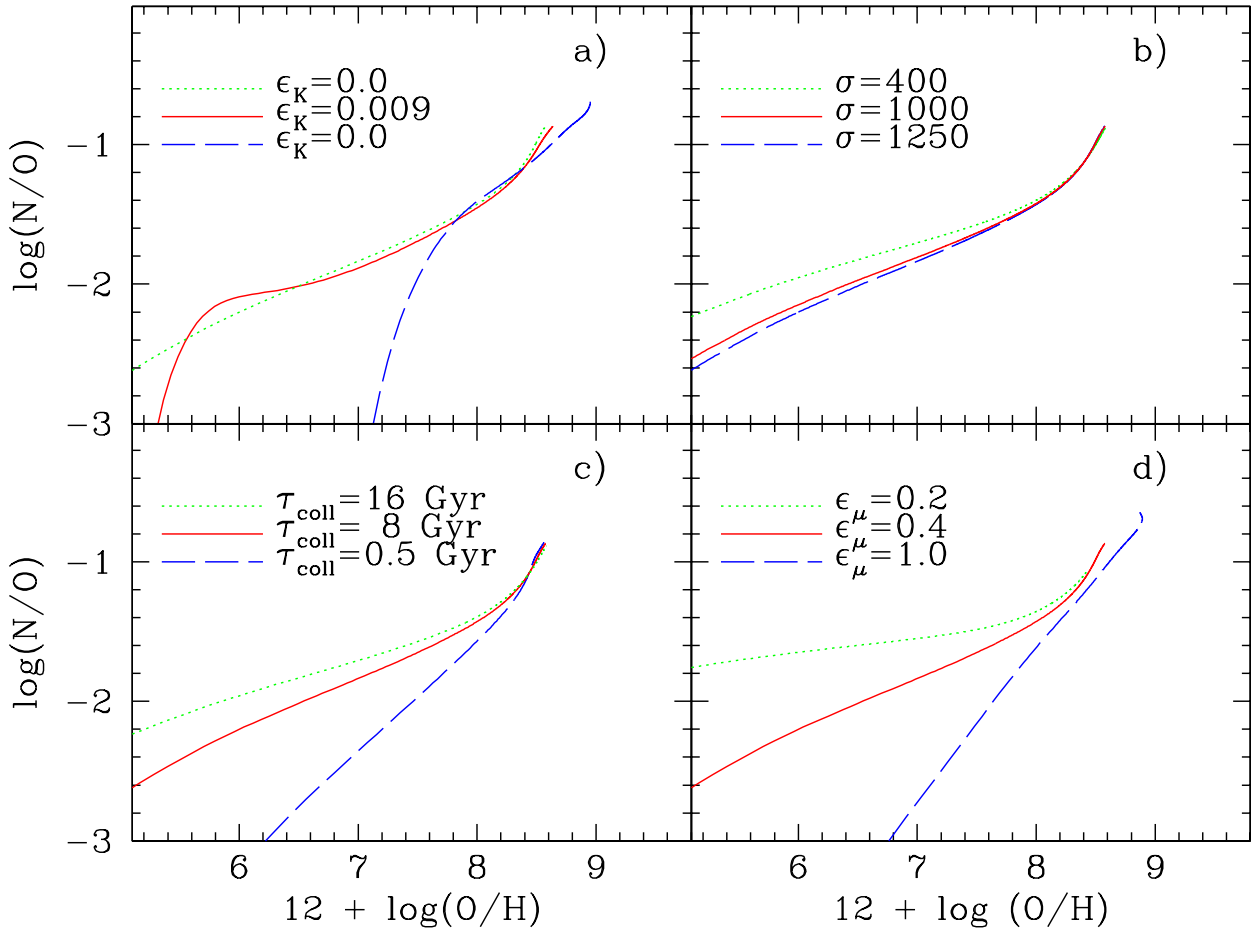}}
\caption{The effect of varying the input parameters of some generic chemical
evolution models on the evolution of the relative abundance 
$\rm log(N/O)$ {\sl vs} the oxygen abundance $\rm 12 + \log(O/H)$: a) The
effect of the star formation of the halo; b) The effect of the
different initial density of gas mass in the region; c) 
the effect of the infall gas rate; d) the effect of the star 
formation efficiencies.}
\label{no_parametros}
\end{figure*}

We, therefore, analyze in detail this influence of these input
parameters \citep[see][for a wide explanation of the multiphase model
and its parameters]{mol05} over the N/O evolution. For that, we have
computed some family of generic models, (that is, they are not models
applicable to MWG) all of them using only a matrix Q
corresponding to one metallicity:

\begin{enumerate}

\item Three models where the star formation rate in the halo,
represented by its efficiency $\epsilon_{K}$, takes three
different values: $\epsilon_{K}$=0, $\epsilon_{K,std}$, and $100
\times \epsilon_{K,std}$, where $\epsilon_{K,std}$ is the value used
in \citet{fer94}, and subsequents multiphase models, for the MWG halo.

\item Three models where there is no star formation in the halo,
$\epsilon_{K}=0$, and where only the density of the initial mass gas
change from a model to the following one.

\item Three models where there is no star formation in the halo,
 $\epsilon_{K}=0$, and where efficiencies to form stars are the same,
 but with a total mass and an infall rate different for each model.

\item Three models where there is no star formation in the halo,
$\epsilon_{K}=0$, and where the total mass and the infall rate is the
same for all models, but where efficiencies to form stars
are different for each one.

\end{enumerate}

The results of these models are shown Fig.~\ref{no_parametros}. In each panel
we show the variations on the N/O track by the change of one parameter as
described.  Its is clear from these results that there are some parameters who
may change the N evolution since the early phases of the evolution. Only in
panel b) N/O shows a similar evolution for all densities. Instead, the
evolution of N/O depends very strongly on the star formation rate (either in
the halo or in the disc). The effect of the halo star formation is
only evident if $\epsilon_{K}$ is so high. If it is low enough the variations
for $12+log(O/H) > 6$ are indistinguishable in the resulting track.

As before, the N/O evolution shows a flatter or a steeper behaviour
depending on the type of scenario or model defined by the parameters
of star formation in the halo, the possible infall rate from the halo
to the disc, and the star formation rate in the disc. We recall that a
strong star formation in the halo in the earliest phases of the
evolution, before stars form in the disc, as it is thought it occurred
in our Galaxy, produces a shift of the N/O track in the plane N/O-O/H
toward high oxygen values and an steep similar to this shown in
Fig.~\ref{evolN} for the model in which the primary nitrogen is
ejected by the lowest mass stars.  On the other hand, if there is no
star formation in the halo, as occurs in the three other panels, or it
is low enough, the abrupt {\sl change of phase} disappears in our
graphs, just because it occurs at lower oxygen abundances than those
shown of the figure. This way the track simulates a smoother behaviour.
Therefore, variations of input parameters and scenarios may explain,
at least partially, the differences found when chemical evolution
models in the literature, using the same set of yields, are compared.

The old version of the Galactic multiphase chemical evolution model
from \cite{fer92,fer94} was computed with the same efficiency to form
stars and molecular clouds for all MWG disc regions, but using
collapse timescales variable with the galactocentric radius of each
radial region, different total masses, and taking into account the
differences in the geometry, that is the effect of the volume over the
densities.  And, hence, the resulting star formation histories show
higher and earlier maximum values at the inner regions of the disc
than at the outer ones. Thus, the results for different radial regions
of the disc already shown a behaviour in agreement with the previous
section: a more {\sl secondary} track for the strong star formation
regions of the inner disc, and less steeper for the outer regions ones
in which the star formation resulted smoother.

\subsection{The effect of the metallicity dependent yields}

The stellar models and their corresponding yields of elements depend
now on metallicity, and, therefore, a set of matrices Q's for each
metallicity is used \footnote{Such as \citet{fer92,pcb98} claim the
$Q_{i,j}$ formalism have some advantages since it compensates the lack
of metallicity dependent yields.  One may think that the use of
matrices Q's may be avoided when this kind of yields are already
available.  However, such as \cite{pcb98} explained: {\sl... stellar
models with different metallicities generally assume solar relative
abundances of the various species within a given Z ; but abundance
ratios are not constant in the course of galactic evolution, nor they
are in the evolution of a chemical model. The matrix links any ejected
species to all its different nucleosynthetic sources, allowing the
model to scale the ejecta with respect to the detailed initial
composition of the star through the Q's}. Therefore, as these authors
we continue using the same formalism.}.

We have computed a Galactic chemical evolution model following the
same prescriptions given in \cite{fer94} and \cite{pcb98} but using
the metallicity dependent yields from \cite{vhoek97} for LIM stars and
WW95 for massive stars.  The resulting model reproduces well the star
formation history and the age-metallicity relation for the Solar
vicinity, as it is shown in GAV05. The radial distributions of diffuse
and molecular gas, star formation rate and oxygen abundance also fit
the observational constraints.

\begin{figure}
\resizebox{\hsize}{!}{\includegraphics[angle=0]{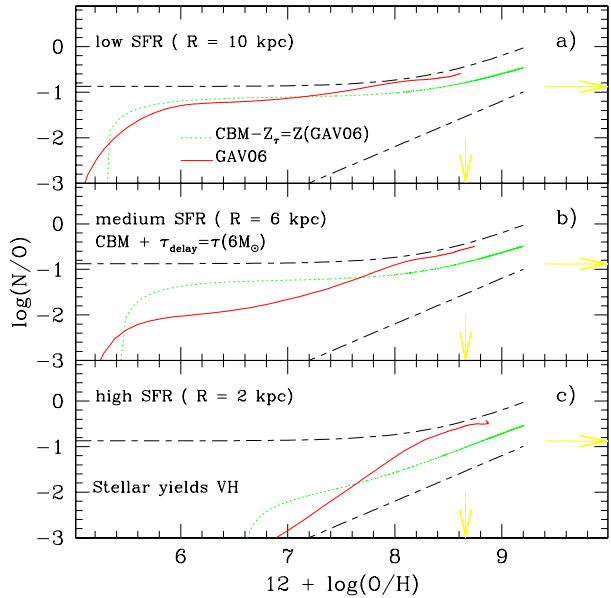}}
\caption{The relation N/O--O/H obtained using the stellar yields for
LIM stars from VH, with the Galactic multiphase chemical evolution
model from \citet{gav06}, for three regions of the Galactic disc
located at a) 10, b) 6 and c) 2 kpc galactocentric radius, solid (red)
lines. Each line is compared with the CBM results (dotted green lines)
computed assuming a delay $Z_{\tau}=Z(t=\tau(6M_{\odot})$ taken from
the corresponding region multiphase model, and using Eq.(8) and (9) as
the dependent on metallicity yields $psN$ and $ppN$. The short-long-dashed 
lines and the arrows have the same meaning than in Fig.~\ref{evolN}.}
\label{no_comp_cb}
\end{figure}

In Fig.~\ref{no_comp_cb} we represent the evolution of $\rm log(N/O)$
{\sl vs} $\rm 12+log(O/H)$ for this model. In each panel we compare
the results for an outer, medium and inner Galactic disc region, as
labelled in the figure,(solid line), with CBM models (dotted lines)
computed assuming that NP appears when stars of 6 $M_{\odot}$ die. The
values for the abundance $Z_{\tau}$, shown in Table~\ref{zmulti}, are
taken from the same multiphase models for $t=\tau(6M_{\odot})$.
 \footnote{Actually, in the multiphase models the maximum star
formation moves toward later times for the outer regions compared with
the early times in which the maximum occurs for the inner
regions. Therefore, the $Z_{\tau}$ might be chosen taken into account
this shift in the time scale respect of the zero time with which we
have selected the values of Table~\ref{zmulti}.}  In this case we have
used the set with $\eta_{AGB}=4$ and $m_{HBB}=0.8 M_{\odot}$ of
metallicity dependent yields from \citet{vhoek97}, which have been
included in the CBM, after a fit with least squares functions, by the
expressions:
\begin{equation}
ppN=1.0E-8\times z_{\tau}^{-0.76}
\end{equation}
and
\begin{equation}
psN=0.29+1598*z_{\tau}
\end{equation}

This way, the values for the yields $ppN$ and $psN$ vary according
with the abundance reached in the ISM when stars forms\footnote{This
method is, in fact, an approximation: if ppN and psN depend on Z, it
is necessary to perform a new integration and the final result is not
equal to Eq. (3). We have checked that this approximation, easier to
understand and to use, is enough good, giving a similar behaviour to
the one obtained with the exact solution.} . For a CBM all stars form
simultaneously, so we have taken for this metallicity the value
$Z_{\tau}$. The resulting yields are given in the two last rows
of Table~\ref{zmulti}. For high metallicity regions, as the inner disc
located at 2 kpc, $Z_{\tau}$ is higher than for a medium star
formation region, as the Solar vicinity, and, therefore, $ppN$ is lower,
and $Z_{N}/Z_{O}=ppN/p$ will be also lower. This means that the
exponential function shown in Fig.~\ref{evolN} increases until a
smaller value. On the other hand, the value psN will be higher for the
inner region than for the Solar Vicinity, and, consequently, the
secondary behaviour due to LIM stars is stronger and more evident for
the region at 2 kpc than for the one located at 6 or 10 kpc.

\begin{table}
\caption{Values of oxygen abundances $Z_{\tau}$ reached when LIM stars
of 6$M_{\odot}$ died in the multiphase models VH from GAV06 and
the corresponding values for psN and ppN}
\label{zmulti}
\begin{tabular}{ccccc}
\hline
Radius & Z$_{\tau}$ &  OH$_{\tau}$ & ppN & psN  \\
(kpc)  &            &              &     &      \\
\noalign{\smallskip}
\hline
\noalign{\smallskip}
2  & 4.7E-5 & 6.60 & 2.01E-5 & 0.365 \\
6  & 3.0E-6 & 5.40 & 1.66E-4 & 0.295 \\
8  & 2.4E-6 & 5.30 & 1.99E-4 & 0.293 \\
10 & 2.0E-6 & 5.24 & 2.23E-4 & 0.292 \\
18 & 1.5E-6 & 5.10 & 2.83E-4 & 0.291 \\
\hline
\end{tabular}
\end{table}

The same behaviour is apparent for the multiphase models.  Thus, as it
is evident in that figure, the metallicity dependent yields change
the expected tracks in the figure N/O-O/H due to variations in the
star formation rate, increasing the existing difference among them,
specially if this star formation rate is strong enough.  When it is very
low, as it occurs in panel a), models give similar results than the
corresponding ones to Fig.3.  When these metallicity dependent yields
are included in a model with strong star formation (panel c), the
evolution changes appreciably compared with the equivalent track obtained
with only one set of yields. This is reasonable: while Z is low, the
model is similar to the one obtained with Q's matrices computed with
yields for only one metallicity. When the oxygen abundances increases,
the stellar production is computed using other set of yields. In
particular, the yield of N increases but the contribution of NS is
relatively higher than before and the NP, that gives the final flat
level of N/O, is smaller. This way, the evolution of N/O in a strong
star formation model behaves like more secondary and the exponential
shape seems smoother.

\begin{figure}
\resizebox{\hsize}{!}{\includegraphics[angle=-90]{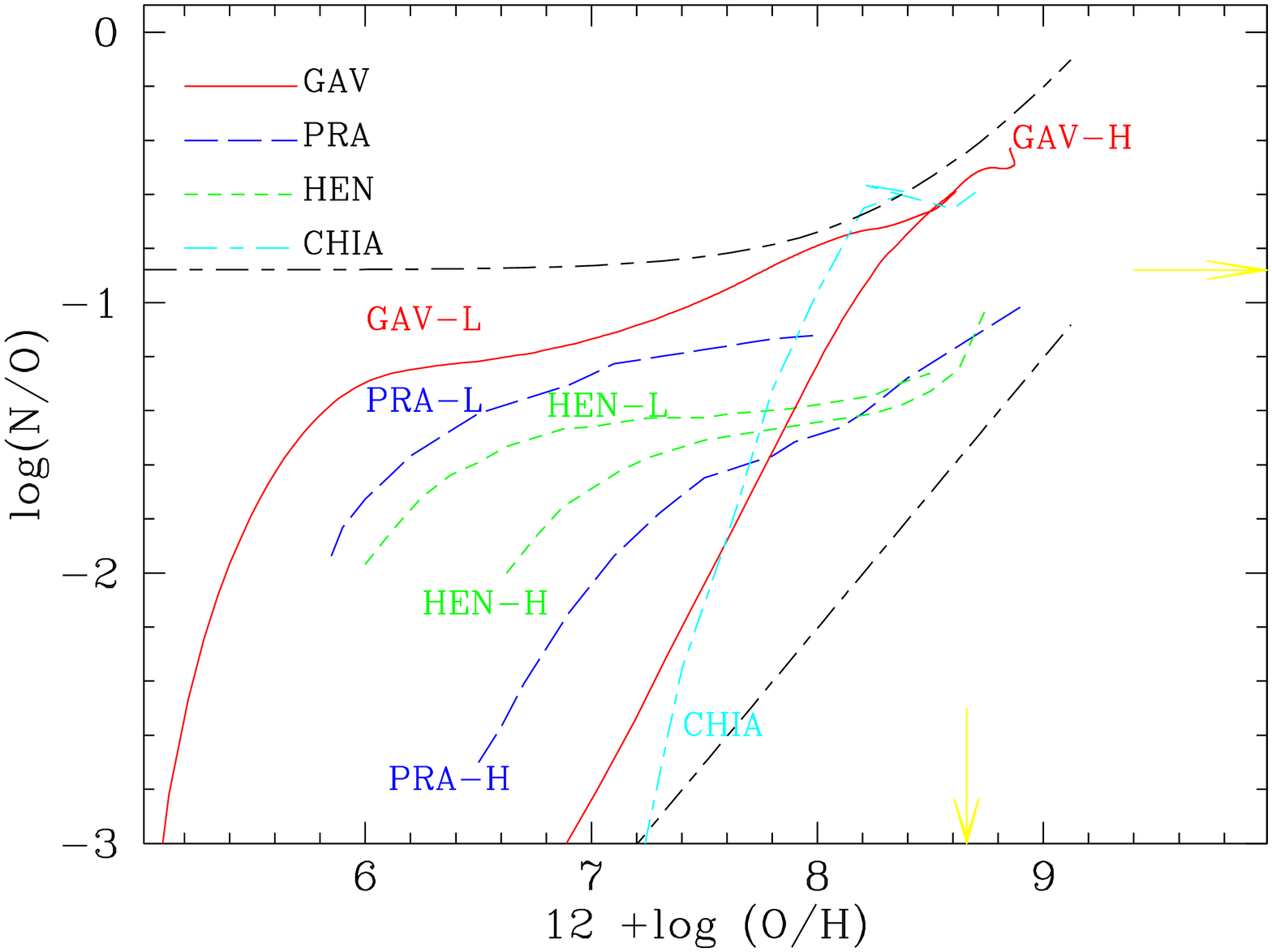}}
\caption{ The evolution of the relative abundance $log(N/O)$ with the
oxygen evolution $12 + log(O/H)$ for two multiphase chemical evolution
models --GAV --,computed using the yields from \citet{vhoek97} and
\citet{woo95} and the mean-lifetimes from \citet{scha92}, compared
with that obtained in models from \citet{hen00,pran03} --HEN and PRA,
respectively, and \citet{chi03-1}--CHIA. The lines coding is given in
the figure. Each model, except CHIA,is marked as L or H following the
star formation rate is low or high. The short-long-dashed lines and
the arrows have the same meaning than in previous figures.}
\label{no_comp}
\end{figure}

\subsection{Comparison with other works}

From the previous section we have found that the track in the plane
N/O-O/H result to be flat when the star formation is low and steep in
the case of strong star formation. This effect is not new since, such
as we have already mentioned, some other works, as
\cite{mouh02,lan03}, have found a similar behaviour, a flattening of
the track in the N/O-O/H plot, when the star formation efficiency
decreases. As an example, we compare in Fig.~\ref{no_comp} our results
with those obtained by other authors.  All of them were computed using
the same set of yields from \citet{woo95} for massive stars and from
\citet{vhoek97} for LIM stars.  Our Galactic models for radial regions located
at galactocentric radii $R=2$ and 10 kpc are represented by the solid
line.  Since each radial region has its own total mass, gas infall
rate, efficiencies, and volume, the star formation history results
different for each one, slow and low in the outer region, and strong
and early in the inner one. Correspondingly, they are marked as GAV-L
and GAV-H, respectively. Other lines represent models from HEN, PRA
and \citet{chi03-1}, marked in the plot. The first two authors give two
models with high (represented with an H) and low (represented with a
L) star formation efficiency. Both authors obtained tracks rather flat
in the plane N/O-O/H when the star formation is low, with a change of
phase at $12+log(O/H) \sim 6-6.3$ while the strong star formation
model show this feature at $12+log(O/H)\sim 7$. The \citet{chi03-1}'s
model shows a steeper behaviour more according with our model for
strong star formation, (probably due to their strong star formation in
the halo).
\begin{figure*}
\resizebox{\hsize}{!}{\includegraphics[angle=0]{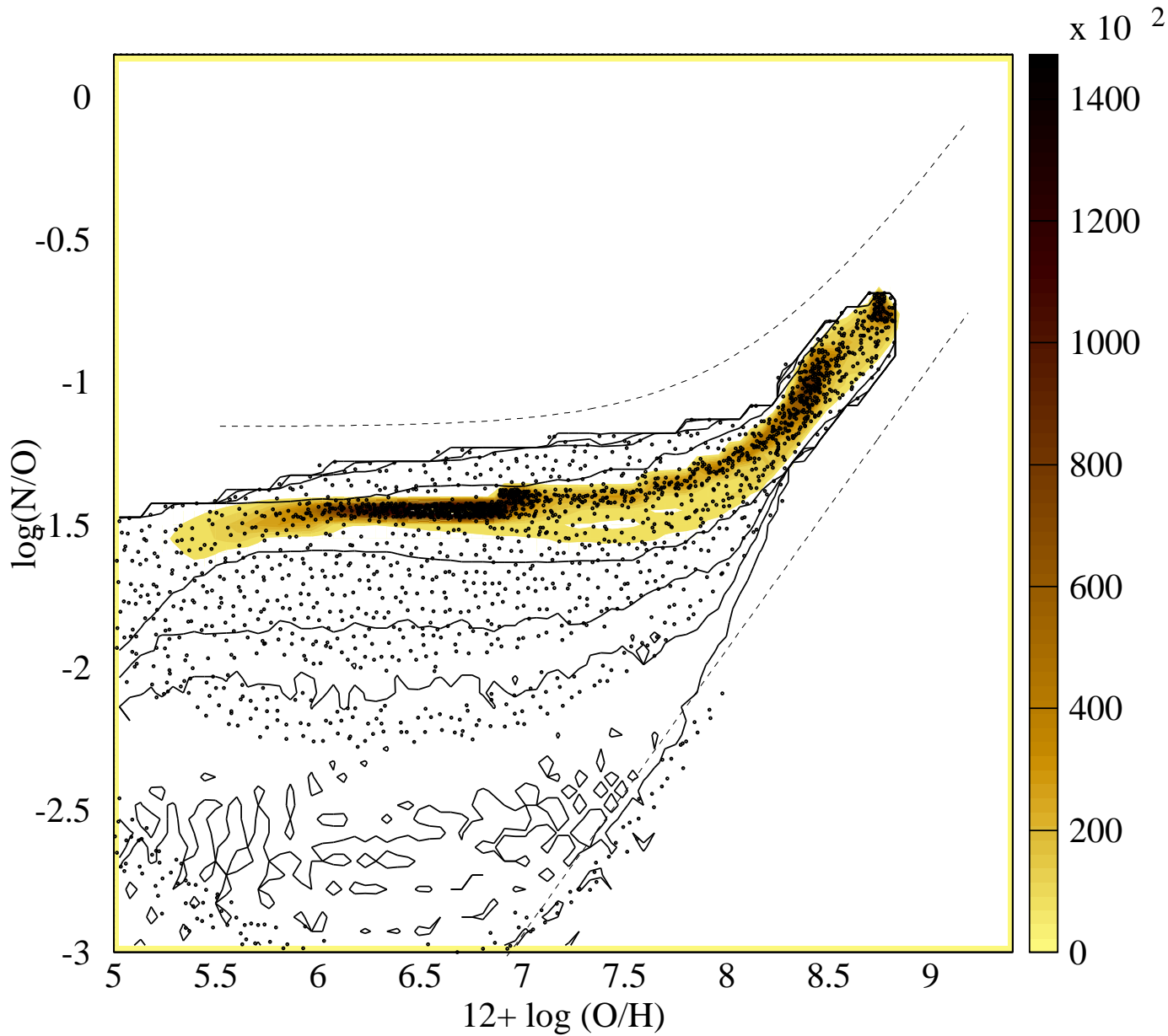}}
\caption{The relation between N/O and O/H for the simulated
galaxies. The complete results for all modeled regions shown as small
dots. Colored contours represent the isolines of number of points
included within them (see relative scale to the right), while solid
lines represent the regions with 0.3, 0.03, 0.012 and 0.003 \%,
respectively, of the total number of points.  The data region is
limited by the two dashed lines.}
\label{no_grid}
\end{figure*}

Is is evident that our models behave similarly to others in what
refers to the variation of N/O tracks with the usual input parameters
of chemical evolution models. We claim that a flat behaviour for the
evolution of N/O {\sl vs} O/H is a usual characteristic of regions or
galaxies where the star formation occurs quietly maintaining a low
value for a long time. The presence of this flat behaviour in the N/O
{\sl vs} O/H plot is not only due to the used yields. 

Since the effect of the star formation is important, it also appears
using any other yield set with primary N ejected by LIM stars.
Obviously, the absolute values of the ratio N/O depend on the yields
set used.  GAV05 presented a set of LIM stars yields which are very
similar to those calculated by \citet{dray03} from accurate stellar
models -- see Fig.5 from GAV06-- but span a wider range in mass and
metallicity as required in chemical evolution models. They were
computed with a similar technique to that used by \citet{vhoek97} but
with updated inputs. The main difference between these two sets of
yields resided in the contribution of primary (or secondary) component
to the total nitrogen yield by LIM stars, which is larger in GAV05 by
factors between 2 and 4 depending on metallicities, even though the
total N integrated yield is larger in \citet{vhoek97} at all
metallicities (see Fig.3 from GAV06).  These yields were included and
successfully calibrated in a GCE model in GAV05 and GAV06, reproducing
the classical observational constraints (Star formation history,
radial distributions of gas, etc.) and obtaining N abundances in
agreement with the MWG stellar data.  In particular, the flat behavior
shown by the metal-poor star data of the Galactic halo results very
well reproduced (see Fig.12 of GAV06).

\section{The grid results: N/O vs O/H for spiral and irregular galaxies}

The grid of chemical evolution models from \citet{mol05}\footnote{See
this work for details about the chemical evolution models,
available in
{http://vizier.u-strasbg.fr/viz-bin/VizieR?-source=J/MNRAS/358/521}}
was parameterized by the initial mass radial distribution. For each
one of our 44 theoretical galaxies, defined by this mass distribution,
10 different sets of efficiencies of molecular cloud formation and
star formation were used, providing 10 evolutionary tracks.  Thus, a
total of 440 models were computed.  For each one, a set of radial
regions, variable according to the size of the theoretical galaxy,
were modeled. Therefore, a large number of possible evolutionary
tracks is finally available to compare with observations. The massive
stars yields from WW and the above cited new yields for LIM stars from
GAV05 have been used for calculating this grid of models.  Even though
that there are still some uncertainties in the AGB phase treatment and
in the quantity of primary nitrogen that LIM stars may create, we
consider that the set of stellar yields from GAV05 has been well
calibrated and, therefore, that it is adequate to our purposes.

We ran models that represent the evolution of different regions of
galaxies, assuming that there is also star formation in the halo (with
a constant efficiency for all models), and that infall rates, initial
masses of gas and efficiencies to form stars are all different from a
modeled galaxy to the following one. All these factors multiply their
effects, changing the resulting star formation history.
Thus, the final results show very different evolutions of the N/O {\sl
vs} O/H as we will show in the following figures.

In Fig.~\ref{no_grid} we show the results for the relative abundance
$\rm \log{(N/O)}$ {\sl vs} the oxygen abundance, $\rm 12 +
\log{(O/H)}$ obtained with an updated grid of models, computed with
the same inputs and hypotheses than \citet{mol05} but using as stellar
mean-lifetimes those computed from \citet{scha92}, instead those ones
from \citet{bur87} used in our previous version. There we show the
whole set of results, \footnote{These updated results are available in
electronic format in {http://wwwae.ciemat.es/\~{}mercedes/grid\_chemev.html}}, 
for all calculated times
\footnote{The time step used for writing results is 0.1 Gyr. Obviously
if we plot results with a smaller time step, the graph will result
more densely populated} and models, as small dots. Over them we have
plot some contours in yellow levels, defined by the scale located at the
right. Furthermore, we draw some other contours as solid lines for
regions with 0.3, 0.03, 0.012 and 0.003 \%, respectively, of the total
number of points.

It is evident that the generic trend obtained with our models using
the new yield set looks very similar to that shown by the data which
lie in the region defined by the dashed lines. The predicted
dispersion is also large, as shown by the observations.

\begin{figure}
\resizebox{\hsize}{!}{\includegraphics[angle=-90]{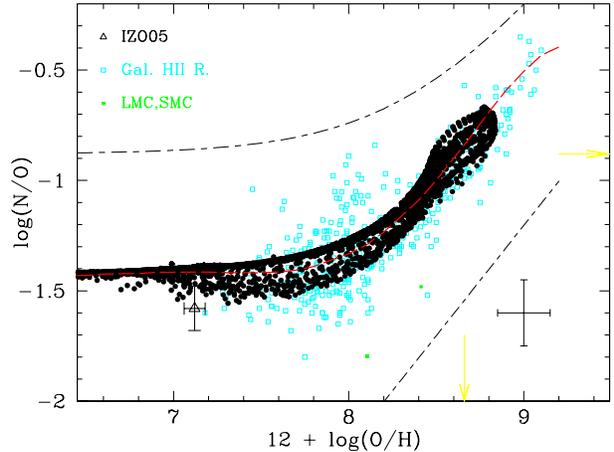}}
\caption{The relative abundance $log(N/O)$ {\sl vs} the oxygen
abundance $12 + log(O/H)$ for the present time as full (black) dots
compared with the data corresponding to Galactic and extragalactic
H{\sc ii} regions, represented as ioen (cyan) squares. The long-dashed
(red) line is the least squares fitting function. The large open
triangle around $12+log(O/H) \sim 7.1$ is the value found by
\citet{izo05} for the lowest-metallicity star-forming galaxy
known. The short-long-dashed lines and the arrows have the same
meaning than in Fig.~\ref{evolN}.}
\label{no_hii}
\end{figure}

In Fig.~\ref{no_hii} we only represent the present time results, as
full points, that we compared with the Galactic and extragalactic
H{\sc ii} regions data, as open squares, taken from references given
in Table~\ref{authors}. Our models may be fitted by a least-squares
polynomial function, shown by a long-dashed line in the same plot:
\begin{eqnarray*}
log(N/O)& = &-1149.31+1115.23x-438.87x^{2}+90.05x^{3}\\
        &   & -10.20x^{4}+ 0.61x^{5} -0.015x^{6}
\end{eqnarray*}
where $x=12+log(O/H)$.

It is evident that our models reproduce adequately the trend of
present day data and, even, a certain dispersion.  There are two zones
where data fall out of the models region. The first one corresponds to
the high metallicity H{\sc ii} ( $Z> Z_{\odot}$) regions, where it is
difficult to estimate directly elemental abundances. In fact, recent
estimations for some of this kind of regions seem to produce lower
oxygen abundances than the old ones, estimated through other empirical
methods \citep{cas02,pil04}.  Therefore, it is probable that those
points must be corrected.  The second region correspond to oxygen
abundances around $12+log(O/H)=8$, just when the parameter $R_{23}$,
usually used to estimated O/H in the low metallicity regions, is
bi-valuated, thus increasing the data errors above the usual error bar
(shown in the plot). On other hand, it is well known that some nearby
star-forming galaxies show evidence of galactic winds and there are a
large number of works \citep{mar04,hens04,rec06,rec06b,rom06}
including this possibility in their evolution models.  We cannot
discard the existence of outflows in the galaxies whose data our
models do not reproduce, and, maybe this is the solution of this
disagreement. However, in general, these winds are important only in
young ($<10^{7}$ yr) starbursts that form many massive stars in a
metal rich environment ($Z> Z_{\odot}$) \citep{vei05}. Recent RX
observations \citep{ras04,ott05,grim05} show evidences of these winds
in dwarf galaxies, but only for those suffering a starburst, which
means SFR $> 5M_{\odot}yr^{-1}$. This high star formation rate is not
always present in dwarf galaxies. On the other hand, the dwarf
irregular galaxies with rotation velocity higher than 30 km.s$^{-1}$
seem able to retain their gas-rich disks \citep{gar02,dek03}. Thus,
the discussion of this subject requires a study for individual
galaxies which fall out of the scope of this work and which we hope
treat in the next future.  In any case the least squares fitting
function, also represented in the figure, has a behaviour in agreement
with the generic trend of data, even at the high-abundance end.

\begin{figure}
\resizebox{\hsize}{!}{\includegraphics[angle=-90]{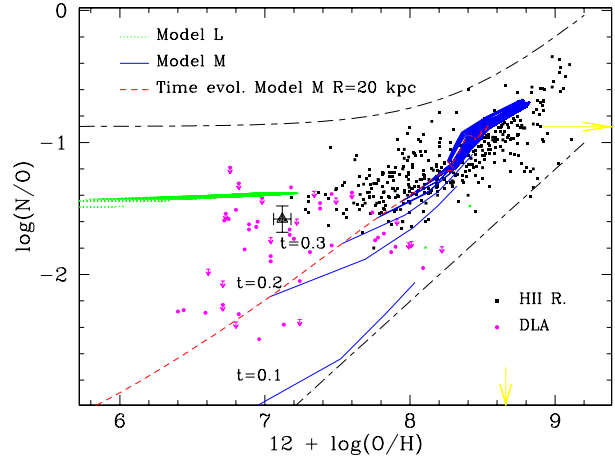}}
\caption{The relation between N/O and O/H for the most evolved and
most massive galaxy, model M, solid (blue) lines), and for the lowest
mass and less evolved galaxy , model L, dotted (green) lines, while the
short-dashed (red) line indicates the time evolution of a region at 
20 kpc of galactocentric distance of model M. The
small full squares are the H{\sc ii} regions data, while the full
(magenta) dots are the DLA objects data.  The large triangle around
$12+log(O/H) \sim 7.1$ is the value found by \citet{izo05} for the
lowest-metallicity star-forming galaxy known. The short-long-dashed 
lines and the (yellow) arrows have the same meaning than in Fig.~\ref{evolN}.}
\label{no_grid2}
\end{figure}

In order to see the effects of the galaxy total mass and/or the
variations of the star formation efficiencies on the computed
abundances, we show in Fig.~\ref{no_grid2} the evolutionary tracks of
N/O abundances for two limiting models: 1) that corresponding to the
most massive galaxy with the highest molecular cloud and star
formation efficiencies --model M, solid lines-- and 2) that
corresponding to the less massive galaxy and with the smallest
efficiencies --model L, dotted lines--.  Each line represent various
radial regions in a same time step, given in the plot in Gyr. We would
like to stress that these lines are not evolutionary tracks in
time. To obtain that, it is necessary to join the points corresponding
to a given radial region of a galaxy at different time steps as we
have done with the short-dashed line.  It is clear from this plot that
the exact mode in which the star formation takes places determines the
evolution of the N/O ratio. In the first type of galaxy, there is an
intense star formation episode, occurring early in time and declining
afterward, simulating an early morphological type bright galaxy.  The
second example shows a low and continuous star formation rate
producing an object similar to a dwarf galaxy.  In the first case N/O
evolves very quickly showing a secondary trend, beginning from $\rm
log(N/O) \sim -2.5$ dex when the oxygen abundance is already $\rm
12+log(O/H)\sim 7$ dex, while in the second N/O keeps a very constant
value of $\rm log(N/O) \sim -1.4 $ dex along the whole evolution, with
oxygen abundances $\rm 12+log(O/H) \leq 7 $.  These two models define
a region similar to the one predicted by \citet[][see their
Fig.A1]{vila93}, but for the opposite reason: the less evolved models
have higher N/O ratios than the ones with strong star formation rates
and smaller time delays.

Thus, the abundances shown by some metal-poor dwarf galaxies, can be
reproduced by models with a continuous star formation rate with low
efficiencies. This is so, even taking into account the lifetimes of
the LIM. In fact a delay in the ejection of N does not necessarily
imply a low value of N/O.  On the one hand because the time required to
reach the N/O value shown by BCD (log N/O $\simeq$ -1.5) might be as
short as $\sim 300-400$ Myr, since intermediate mass stars of
4--8$M_{\odot}$ (which contribute to the primary N), have
mean-lifetimes in the range $ 50 \leq \tau \leq 200$ Myr. On the other
hand, and more important, because low oxygen abundances does not imply
necessarily a short evolutionary time since with a low star formation
it is possible to maintain a low value of O/H for a long time.

These results are only partially due to the dependence with
metallicity of the yields we use.  The integrated ratio of primary to
total N yields $\rm ^{14}N_{P}/^{14}N$ \citep[see Fig.3 from][]{gav06}
has a value $\sim 0.20$ for $Z_{\odot}$ \citep[slightly smaller than
the one estimated, $\sim 1/3$, by ][ as necessary to reproduce the
data]{allo79} and it increases up to 0.6 for low metallicities. This
contribution of primary N, larger for low metallicity stars than for
high metallicity ones, is important since it allows to us to obtain
tracks in the plane N/O {\sl vs} O/H flatter than the ones predicted
with a constant contribution of the primary N. We must to do clear,
however, that if we eliminate the metallicity dependence from the
yields, we still obtain different tracks for regions with different
star formation histories, as we have shown in previous sections. Thus,
the selection of the set of yields is not essential to obtain these
differences among tracks, although the shown effect increases when
metallicity-dependent yields are included in the models.  Furthermore,
the absolute level of observations and the fine tuning of the observed
shape in the plane N/O-O/H for the present time data is only
reproduced if the stellar yields, in turn, give the right level of
primary nitrogen and have the adequate dependence on metallicity. The
chemical evolution models must also be well calibrated in order to
predict star formation histories able to produce abundances in
agreement with data. Therefore, the adequate selection of metallicity
dependent LIM star yields, as those from GAV05, joined to the use of
accurate chemical evolution models, as those from \cite{mol05}, allows
to obtain abundances for N and O which reproduce the complete set of
data in the plane N/O {\sl vs} O/H.

\begin{figure}
\resizebox{\hsize}{!}{\includegraphics[angle=-90]{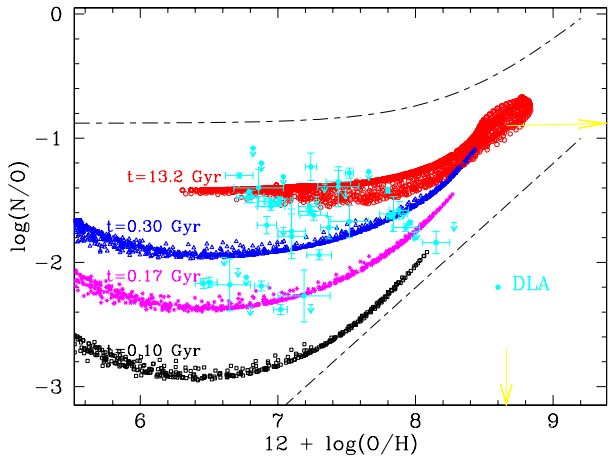}}
\caption{The relation N/O -- O/H for different evolutionary times as
marked in the figure. Full (magenta) dots correspond to DLA objects.}
\label{no_grid3}
\end{figure}

Finally, in order to explore the time evolution of the modeled
abundances, we show in Fig.~\ref{no_grid3} the results for four
different time steps: $\rm t=0.1$, $\rm t=0.17$, $\rm t=0.30$, and
$\rm t=13.2$ Gyr.  These epochs would correspond to redshifts
z$\sim$3.8, 3.7, 3.5, and 0, respectively, for a cosmology with
H$_{0}=71$, $\Omega_{M}=0.30$, $\Omega_{\Lambda}=0.70$, if the
formation of these spiral and irregular galaxies occurred at a
redshift z$\sim 4$. A large gap appears between the results for each
one of our time steps. In particular, our models predict a feature
similar to the so-called {\sl second plateau} that \citet{cen03} claim
to exist at a metallicity around $\rm 12+log(O/H)\sim 7$, which
appears at a level of $\rm \log{(N/O)}\sim -2.3$ for $\rm 12+log(O/H)
\sim 6.5-7$, while most points appear at a higher level of N/O ($ \ge
-1.8$) for a similarly low O/H abundance. In fact, no gap is apparent
between models at 1.1 and 13.2 Gyr, thus making it difficult to
discriminate objects at a redshift up to $z < 2.5$ from those at
redshift z=0 in the N/O-O/H plane, as it is actually the case.  The
abundances predicted for galaxies at redshift $z \ge 3.$ are far
enough from the rest of the points in the plot so as to disentangle
them from a given data sample.

Thus, the abundance data from DLA objects are easily reproduced by
models of massive galaxies with strong star formation in the early
stages of their evolution. This evolution is so rapid in those
galaxies that 300 Myr after the beginning of the formation of stars
the value of log(N/O) increases from -2.5 to -1.6 with oxygen
abundances $\rm 12 +log(O/H)= 7 - 8.5 $.  Therefore, it should be
possible to find objects in this zone of the N/O {\sl vs} O/H plot,
although, given the shortness of this phase, the number of these
objects would be small.  We note that a level of $\rm \log{(N/O)}\sim
-1.6$ dex is easily reached in such a short time, in galactic
evolutionary terms, as $t\sim 300 Myr$, even with a primary
contribution proceeding from LIM stars.

\section{Conclusions}

The evolutionary track followed by a given region or galaxy in the N/O
{\sl vs} O/H plane depends strongly on the star formation history 
of the region. Strong bursting star formation histories would produce high
oxygen abundances soon and, hence, an early secondary behaviour, thus
reproducing most of the spiral H{\sc ii} region observations. On the
contrary, a low and continuous star formation rate keeps the oxygen
abundance low for a long time; thus a large quantity of primary N may
be ejected reproducing the flat slope shown by dwarf galaxy data.

The resulting dispersion of our realizations, similar to the one
observed, appears as a consequence of the different star formation
histories in each modeled region and in each galaxy. These are
obtained by combining various star and molecular cloud formation
efficiencies and collapse time scales.

The stellar yields for LIM also play a role: if they have the adequate
contribution of primary N depending on metallicity, it is easy to
reproduce adequately the generic trend of the observational points in
the N/O {\sl vs} O/H plane, in particular a flat slope for low
metallicity regions and, simultaneously, lower values of N/O at similar
oxygen abundances for regions or galaxies where the star formation is
strong and occurred early in the evolution.

The combination of the new yield set from \citet{gav05} and the
self-consistent chemical evolution models by \citet{mol05} are able to
reproduce the data corresponding to H{\sc ii} regions in different
galaxies and Galactic stars of different ages. This arises naturally
just by assuming different input parameters (mostly star formation
efficiencies and infall rates) for the different galaxies and regions
in the same generic model, without the need to appeal to {\sl ad hoc}
models to reproduce each kind of object.

They may also to reproduce the existing data on galaxies at 0 $\leq$ z
$\leq$ 4, and predict the existence of a {\sl second plateau}, found
in the N/O {\sl vs} O/H relationship for $\rm log(N/O) \sim -2.5$ and
galaxies with $z\ge 3$. Our models predict that objects at redshifts
$z \le 2.5$ can not be discriminated from $z=0$ objects in the
log(N/O) {\sl vs} log(O/H) plane; however, for $z\ge 3$ objects, this
discrimination is possible.  This characteristic is a
consequence of the rapid passage of model results by this zone of the
plane, when the time evolution is considered, since 300 Myr after the
beginning of the formation of stars, the value log(N/O) increases from
-2.5 to -1.6 with oxygen abundances $\rm 12+log(O/H)=7-8.5$

\section*{Acknowledgments}
This work has been partially supported by the Spanish PNAYA project
AYA2004--8260-C03. We acknowledge their comments to R.B.C Henry and
L. Pilyugin who have kindly revised part of this manuscript.
We thank an anonymous referee for the useful comments and suggestions
that have improved this paper.

\bibliographystyle{mn2e.bst} 
\bibliography{bibliografia}
\end{document}